\documentclass[fleqn,usenatbib]{mnras}

\usepackage{newtxtext,newtxmath}

\usepackage[T1]{fontenc}
\usepackage{multirow}
\usepackage{rotating}
\usepackage{lscape}

\DeclareRobustCommand{\VAN}[3]{#2}
\let\VANthebibliography\thebibliography
\def\thebibliography{\DeclareRobustCommand{\VAN}[3]{##3}\VANthebibliography}

\usepackage{graphicx}	
\usepackage{amsmath}	

\title[MeerKAT radio observations of Cen~X--4]{MeerKAT radio observations of the neutron star low-mass X-ray binary Cen~X--4 at low accretion rates}

\author[Van den Eijnden et al.]{J. van den Eijnden,$^{1}$ R. Fender,$^{1,2}$ J. C. A. Miller-Jones,$^{3}$ T. D. Russell,$^{4}$ P. Saikia,$^{5}$  G. R. Sivakoff,$^{6}$ \newauthor and F. Carotenuto,$^{1}$
\\
$^{1}$Department of Physics, Astrophysics, University of Oxford, Denys Wilkinson Building, Keble Road, Oxford OX1 3RH, UK\\
$^{2}$Department of Astronomy, University of Cape Town, Private Bag X3, Rondebosch 7701, South Africa\\
$^{3}$International Centre for Radio Astronomy Research, Curtin University, GPO Box U1987, Perth, WA 6845, Australia\\
$^{4}$INAF, Istituto di Astrofisica Spaziale e Fisica Cosmica, Via U. La Malfa 153, I-90146 Palermo, Italy\\
$^{5}$Center for Astro, Particle and Planetary Physics (CAP$^3$), New York University Abu Dhabi, PO Box 129188, Abu Dhabi, UAE\\
$^{6}$Department of Physics, CCIS 4-181, University of Alberta, Edmonton, AB, T6G 2E1, Canada\\
}

\date{Accepted XXX. Received YYY; in original form ZZZ}

\pubyear{2021}

\begin{document}
\label{firstpage}
\pagerange{\pageref{firstpage}--\pageref{lastpage}}
\maketitle

\begin{abstract}
Centaurus~X--4 (Cen~X--4) is a relatively nearby neutron star low-mass X-ray binary that showed outbursts in 1969 and 1979, but has not shown a full outburst since. Due to its proximity and sustained period of quiescence, it is a prime target to study the coupling between accretion and jet ejection in quiescent neutron star low-mass X-ray binaries. Here, we present four MeerKAT radio observations at $1.3$ GHz of Cen~X--4, combined with \textit{NICER} and \textit{Swift} X-ray monitoring. During the first and most sensitive observation, Cen~X--4 was in a fully quiescent X-ray state. The three later and shorter observations targeted a brief period of faint X-ray activity in January 2021, which has been referred to as a `mis-fired' outburst. Cen~X--4 is not detected in any of the four MeerKAT observations. We place these radio non-detections on the X-ray -- radio luminosity diagram, improving the constraints on the correlation between the two luminosities from earlier quiescent radio studies. We confirm that Cen~X--4 is radio fainter than the transitional milli-second pulsar PSR J1023+0038 at the same X-ray luminosity. We discuss the radio behaviour of accreting neutron stars at low X-ray luminosity more generally and finally comment on future observing campaigns.

\end{abstract}


\section{Introduction}

Low-mass X-ray binaries (LMXBs), wherein a black hole (BH) or neutron star (NS) accretes from a binary companion star, are routinely used to study the connection between accretion and the launch of outflows. Such outflows can take two forms, namely highly-collimated and often relativistic outflows from the inner accretion flow called jets, and slower, more massive disk winds with a wider opening angle and a range of launch radii. A large fraction of LMXBs are transient sources, spending most of their lifetimes in quiescent states where little to no accretion takes place. Intermittently, these transient LMXBs show outbursts, where the accretion disk transitions into a hot, ionized state and the accretion rate increases by orders of magnitude. Alternatively, persistently-accreting LMXBs do not show outburst-quiescence cycles, and their accretion disk instead remains in the ionized outburst state \citep[see e.g.][for reviews]{done07,gilfanov10}.

LMXBs in outburst can reside in different spectral-timing states \citep{fender04}: the system rises in accretion rate in the hard state, where its X-ray emission is highly variable and dominated by the Comptonized emission from an optically-thin population of hot electrons often called the corona. Via intermediate spectral-timing states, it then often transitions into the soft state, characterized by weak X-ray variability and emission from an optically-thick, geometrically-thin disk. While it may transition to a high Eddington accretion state or make multiple transitions between the hard and soft state, during the outburst decay, it transitions from the soft into the hard state at lower X-ray luminosities than the hard-to-soft transition \citep{maccarone2004}. Finally the outburst sinks back into quiescence. The spectral-timing states of NSs are historically named differently, with the main classification into atoll and Z-sources based on the shape of the tracks they display in their X-ray color-color diagrams. Within both the atoll and Z-classifications, several sub-states have been identified \citep[see e.g.][for details]{hasinger89,vanderklis06}. The discovery of a NS LMXBs transitioning between the atoll and Z-source classifications showed that mass-accretion rate fundamentally underlies the difference between the classes, with the latter at the highest accretion rates \citep{homan07}. Despite these different historical spectral-timing classifications between NS and BH systems, several authors have identified equivalencies between the states across both source types \cite[e.g.][]{migliari06,munozdarias14}.

These LMXB states are intricately linked to their outflow properties. In BH systems, the hard state is associated with the continuous launch of a compact jet, followed by the launch of discrete ejecta during the hard-to-soft transitional states, while the compact jet in quenched in the soft state \citep[e.g.][]{fender04,russell2019_1535}. In atolls, which can (but do not always) change between Comptonization-dominated (`hard', island state) and thermal-dominated (`soft' banana state) states, compact jets are observed routinely in the former \citep{migliari03, migliari06, gusinskaia20}. However, whether jet quenching in the latter state always occurs remains unclear \citep{rutledge98,migliari04,millerjones10,fender16,gusinskaia17,vandeneijnden2021}. The Z-sources show compact jets and discrete ejecta, depending on the exact branch of their Z-track \citep{migliari06}. Disk winds were initially observed in the X-ray band, where they show up in BH soft states and are easiest to observe in high inclination systems \citep{ponti12}. However, disk winds are also observed in the optical \citep[e.g.][]{munozdarias16} and IR band \citep[e.g.][]{sanchez2020}: optical winds are observed in the hard state, while IR winds are seen across the outburst. Recently, the first UV wind detection was reported in a NS LMXB \citep{castrosegura2022}. Finally, outburst light curve modelling also suggests strong outflows \citep{tetarenko18_winds}.

The connection between the accretion flow and compact jets is routinely studied in the X-ray -- radio luminosity ($L_X$--$L_R$) diagram. Here, the former traces the accretion luminosity and is a proxy for accretion rate, while the latter traces the jet and its brightness. Compact BH jets show an $L_X$--$L_R$ correlation across $\sim$eight orders of magnitude in X-ray luminosity \citep{hannikainen98,corbel00,corbel03,gallo2006}; a subset of sources follows a radio-bright correlation with a power law slope of $\beta \approx 0.6$ (where $L_R \propto L_X^{\beta}$, while others follow a steeper correlation with $\beta \gtrsim 1$ at high X-ray luminosities, before re-joining the other track around $L_X \approx 10^{35}$ erg/s \citep[e.g.][although discussion exists regarding the statistical robustness of this separation; e.g. \citealt{gallo18}]{soleri11,coriat11,carotenuto2021}. The situation for NS LMXBs is even more complex. While, as a sample, NS LMXBs are $\sim 22$ times radio-fainter than BH systems, the individual NS systems do not appear to follow a single correlation; significant scatter exists both between sources and between outbursts of the same source. Furthermore, due to their radio faintness, few sources have been monitored over a large range of X-ray luminosity --- particularly below $L_X \approx 10^{35}$ erg/s, few NS LMXBs have been detected in radio \citep[e.g.][]{gusinskaia20,tudor17}. For the sample of NS LMXBs, however, a power law slope of $\beta \approx 0.4--0.5$ has been measured \citep{gallo18}, which is similar to BHs. Different radiative efficiencies may be expected for the two types of LMXBs --- BHs can advect a fraction of the liberated gravitational energy across the event horizon, while the presence of a NS surface implies that all this liberated energy should, eventually, be radiated. Therefore, a similarity in the $L_X$--$L_R$ coupling of the two source classes, which depends on this radiative efficiency, is surprising.

While BH systems have been radio-detected in quiescent states down to $L_X < 10^{31}$ erg/s \citep{gallo2006,dincer18}, no NS LMXB radio detections have been obtained below $L_X = 10^{34}$ erg/s.\footnote{Here, we ignore the transitional millisecond pulsars, which we will discuss in Section \ref{sec:disc}.} Moreover, few radio upper limits exist in this X-ray regime. Therefore, the correlation between the accretion flow and compact jets is poorly known for weakly-accreting and quiescent NS LMXBs, and either radio detections or deeper upper limits in this regime are necessary to advance our understanding. In this work, we present a dedicated radio and X-ray campaign of the NS LMXB Centaurus~X--4 (Cen~X--4) to constrain the low-$L_X$ inflow-outflow coupling.

Cen~X--4 is a close-by transient NS LMXB, located at a relatively close-by distance: \citet{chevalier1989} determine $d=1.2\pm0.3$ kpc, while recent Gaia measurements imply a slightly larger distance (see Section \ref{sec:disc}). It was discovered originally in outburst in 1969 \citep{conner1969} and showed a second outburst in 1979 \citep{kaluzienski1980}. Observations with the Very Large Array detected its radio counterpart and monitored it during the 1979 outburst \citep{hjellming1979,hjellming1988}. Searches for radio emission during quiescence yielded no detections, despite its small distance \citep{kulkarni1992,tudor17}. However, the limit found by \citet{tudor17} provides the deepest radio constraint on a quiescent NS LMXB and therefore provides the best low-$L_X$ anchor to the NS $L_X$--$L_R$ relation to date. In January 2021, long-term optical monitoring of Cen~X--4 in the XB-NEWS program \citep{waterval2020} revealed the onset of activity in all optical bands \citep{saikia2021}. X-ray follow up observations confirmed the activity \citep{vandeneijnden2021_atel1}. However, the source did not enter a full outburst phase, but instead returned to quiescent levels at all wavelengths two weeks later \citep{vandeneijnden2021_atel2}. This behaviour has been referred to as a `mis-fired' outburst, caused either by an inside-out heating front stalled by low levels of irradiation in the outer accretion disk, or by a local thermal-viscous instability in the disk \citep{baglio2022}. In this paper, we present MeerKAT radio observations of Cen~X--4 taken in its quiescent state in September 2020 and during the brief period of activity in January 2021.

\section{Observations and data reduction}

\begin{table}
\caption{Summary of the MeerKAT radio observations of Cen X-4. The Start MJD column refers to the first on-target scan, and the On-source time does not include setup or calibration scans. The radio luminosity $L_R$ is calculated assuming a distance of $1.2$ kpc, and limits are quoted at $3$-$\sigma$ significance.}
\label{tab:radio}
\begin{tabular}{lllll}
\hline
 & Start MJD & On-source time & RMS sensitivity & $5$-GHz $L_R$ \\
& & & [$\mu$Jy/bm] & [erg/s] \\ \hline
1 & 59118.54 & $4$ hour & $4.3$ & $<1.1\times10^{26}$ \\
2 & 59221.43 & $15$ min & $23$ & $<5.9\times10^{26}$ \\
3 & 59223.36 & $15$ min & $18$ & $<4.7\times10^{26}$ \\
4 & 59230.32 & $15$ min & $16$ & $<4.1\times10^{26}$ \\
\hline
\end{tabular}
\end{table}

\subsection{Radio: MeerKAT}

We performed a total of four observations of the field surrounding Cen~X--4 (J2000 14$^{\rm h}$58$^{\rm m}$21.935$^{\rm s}$, $-$31$^{\circ}$40$^{'}$07.52$^{''}$) with MeerKAT. The first observation was taken on 26 September 2020 (capture block 1601122564), while Cen~X--4 was in its usual quiescent state, with a longer, $4.5$-hour observation time including overheads ($\sim 4$ hour on target). The three remaining observations were taken during the brief period of activity in January 2021, specifically on 7, 9, and 16 January 2021 (capture blocks 1610014143, 1610180483, 1610782322, respectively). Those latter three observations were all shorter, providing $15$ minutes on-target observing time. All observations were taken as part of the ThunderKAT Large Survey Program \citep{fender_MKAT}, which monitors active transient X-ray binaries on a weekly basis and performs additional dedicated observations of individual X-ray binaries, such as the quiescent observation of Cen~X--4. Observational details are also summarized in Table \ref{tab:radio}.

All four observations were performed using the L-band receivers, providing an observing band between $856$ and $1712$ MHz, for a central frequency of $1.3$ GHz. The data were collected in 32k-mode (i.e. 32768 frequency channels) and standard integration time of 8 seconds. The (standard) primary and nearby secondary calibrator sources were J1939-6342 and J1501-3918, respectively. The former was observed at the start of each observation for $5$ minutes. $2$-minute scans of the latter source bookended the target scan during the three short observations, while such scans were performed every $30$ minutes during the longer September 2020 observation.

We used the \textsc{OxKAT} suite of analysis scripts \citep{heywood2020}\footnote{\url{https://github.com/IanHeywood/oxkat}} to perform flagging, calibration, and imaging of the observations. Through \textsc{OxKAT} we used the \textsc{common astronomy software application} \citep[\textsc{casa};][]{mcmullin07} to perform initial data averaging, flagging, and calibration. Afterwards, we applied the bandpass, flux-scale, complex gain, and delay corrections to the target scans, and split of the target data, before flagging and imaging these data using the \textsc{tricolour} and \textsc{wsclean} \citep{offringa2014} packages, respectively. Next, we applied self-calibration and re-imaged the field. Finally, (as no source was detected at the position of Cen X-4), we placed upper limits on the flux density at three times the RMS sensitivity over a region covering the source position that was also devoid of detected point sources in the longer observation.

\subsection{X-rays: \textit{Swift} and \textit{NICER}}

During the time frame of the MeerKAT observations, the \textit{Neil Gehrels Swift Observatory} (hereafter \textit{Swift}) X-ray Telescope (XRT) observed Cen~X--4 seventeen times. On 26 September 2020 (ObsID 00088937006), a coordinated observation was performed simultaneously with the deep MeerKAT observation, for a total exposure of $\sim 3$ ks in Photon-Counting (PC) mode. Later, during the brief period of activity around January 2021, \textit{Swift}/XRT observed a further sixteen times in PC mode between 28 December 2020 and 4 February 2021 with shorter exposures ranging from $\sim 0.2$ to $\sim 1.0$ ks. To monitor the profile of the X-ray activity, we extracted count rates for all observations using the Online Data Products Pipeline \citep{evans07}\footnote{\url{https://www.swift.ac.uk/user\_objects/}}.

To measure the X-ray flux evolution during the observations, we also extracted the X-ray spectrum for each observation. We downloaded the raw data from the \textsc{heasarc} and used the \textsc{run\_xrtpipe.pl} script to perform standard data reduction. We then created source and background regions by defining a circular source region with a $59$ arcsec diameter \citep[25 pixels;][]{evans07} centered on the brightest target pixel and an annular background region with inner and outer diameters of 59 and 400 arcsecond, respectively. Due to the low flux of Cen~X--4 and the short \textit{Swift} exposures, five observations did not detect sufficient counts for a spectral analysis (i.e. between 2 and 17 counts in the source region). Therefore, we only extracted source and background spectra from the remaining twelve observations, using \textsc{xselect}. We then created the observation-specific \textsc{arf}-response files using the \textsc{xrtmkarf} tool and downloaded the appropriate \textsc{rmf}-response file, \textsc{swxpc0to12s6\_20130101v014.rmf}, from the \textsc{caldb}. Finally, we rebinned all twelve spectra to at least $1$ count per bin using \textsc{grppha}. In all steps we used \textsc{HEASoft} v6.29. We also used \textit{Swift's} Online Data Products Pipeline to extract the X-ray spectra for the same observations, to re-perform our spectral analysis and confirm that the results are consistent.

\textit{Swift}/XRT did not observe within 24 hours from the 9 January 2021 MeerKAT observation. However, the \textit{Neutron Star Interior ExploreR} (\textit{NICER}) observed with a $\sim 0.53$ ks cumulative\footnote{\textit{NICER} ObsIDs typically combine data from multiple exposures on the same day, whose individual durations are short due to the complex visibility of targets as seen from the International Space Station.} exposure on that day (ObsID 3652010901). \textit{NICER} is a single-pixel instrument with a lower angular resolution than \textit{Swift}/XRT (particularly in PC mode), a softer response, and higher background count rate. Therefore, \textit{NICER} spectra should be examined carefully when observing faint sources such as Cen~X--4, for instance to determine the energy range where the source dominates over the background. Hence, we only extract the \textit{NICER} spectrum when no \textit{Swift} data is available. To extract the \textit{NICER} spectrum, we downloaded the \textsc{heasarc} datafile and ran the \textsc{nicerl2} tool to re-apply the level-2 calibration using the latest version of the \textsc{caldb}. We then used \textsc{xselect} to extract the target spectrum, without specific energy cuts, as we checked the energy range where the source dominates explicitly before spectral fitting. We then created response files using the \textsc{nicerrmf} and \textsc{nicerarf} tools, before calculating the background spectrum using the \textsc{nicer\_bkg\_estimator} tool\footnote{See \url{https://heasarc.gsfc.nasa.gov/docs/nicer/tools/nicer\_bkg\_est\_tools.html}}. We finally rebinned the spectrum to 20 counts per bin.

All X-ray spectral fits, for both \textit{Swift} and \textit{NICER} observations, were performed in \textsc{xspec} v12.12.0 \citep{arnaud96}, assuming interstellar abundances from \citet{wilms00} and cross-sections from \citet{verner96}. We included the interstellar absorption via the \textsc{tbabs} model and used the convolution model \textsc{cflux} whenever we calculated fluxes from the fitted models. We calculated and report all errors at the $1$-$\sigma$ level. Due to the low X-ray count rates, we applied C-statistics \citep{cash1979} for all X-ray spectral fitting.

\section{Results}

\subsection{Radio observations}

\begin{figure*}
    \centering
    \includegraphics[width=\textwidth]{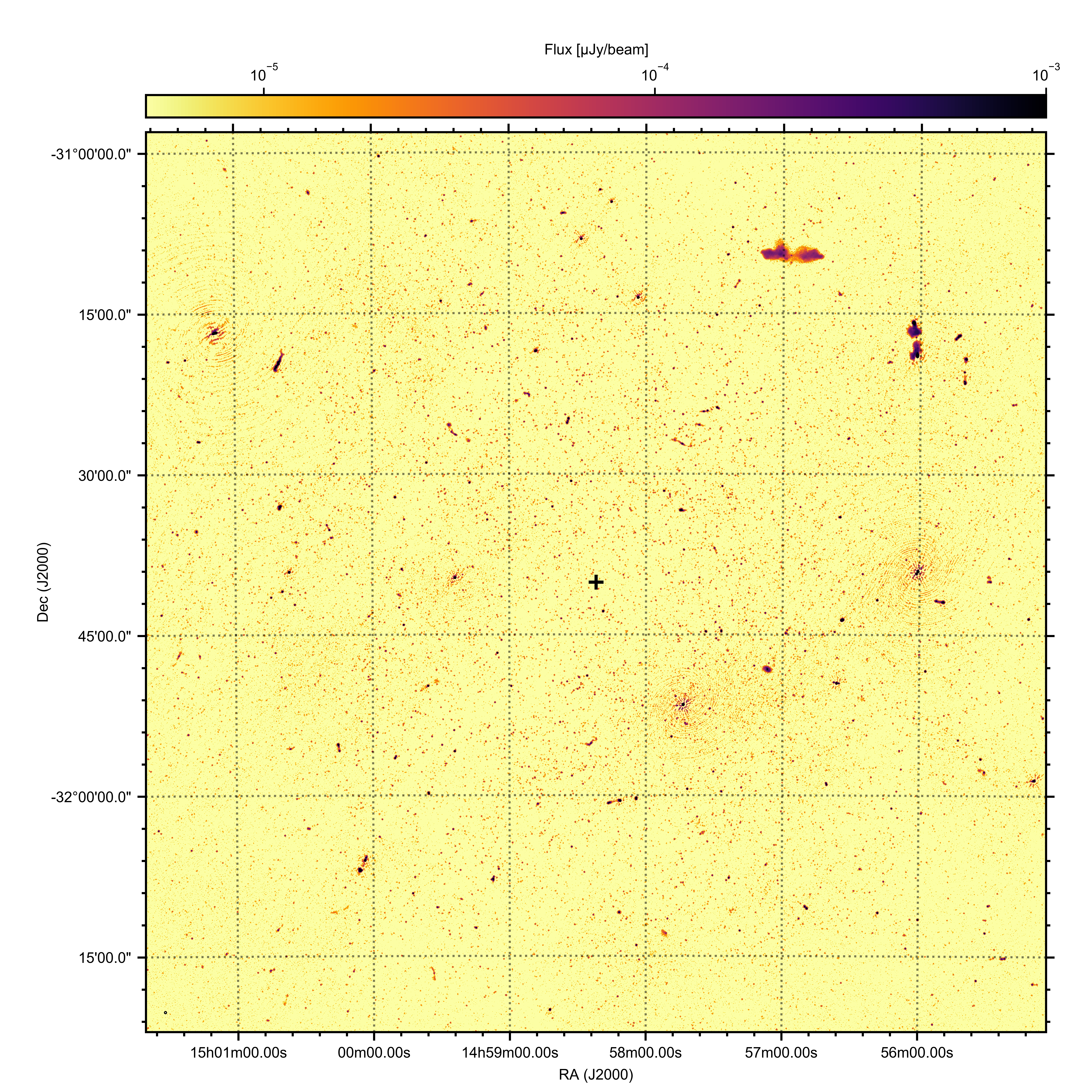}
    \caption{The large-scale $1.3$ GHz MeerKAT view of the field surrounding the position of Cen~X--4, shown by the black cross. The RMS noise of the image is $4.4$ $\mu$Jy/beam, obtained in the $4.5$-hour observation on 26 September 2020. The beam is shown in the lower left corner. A large number of unresolved point sources and extended objects, are visible throughout the $1.4\times1.4$ degree field. A zoom around the position of Cen~X--4 is shown in Figure \ref{fig:image_zoom}, confirming the absence of a MeerKAT radio detection.}
    \label{fig:image_largeFOV}
\end{figure*}

\begin{figure}
    \centering
   \includegraphics[width=\columnwidth]{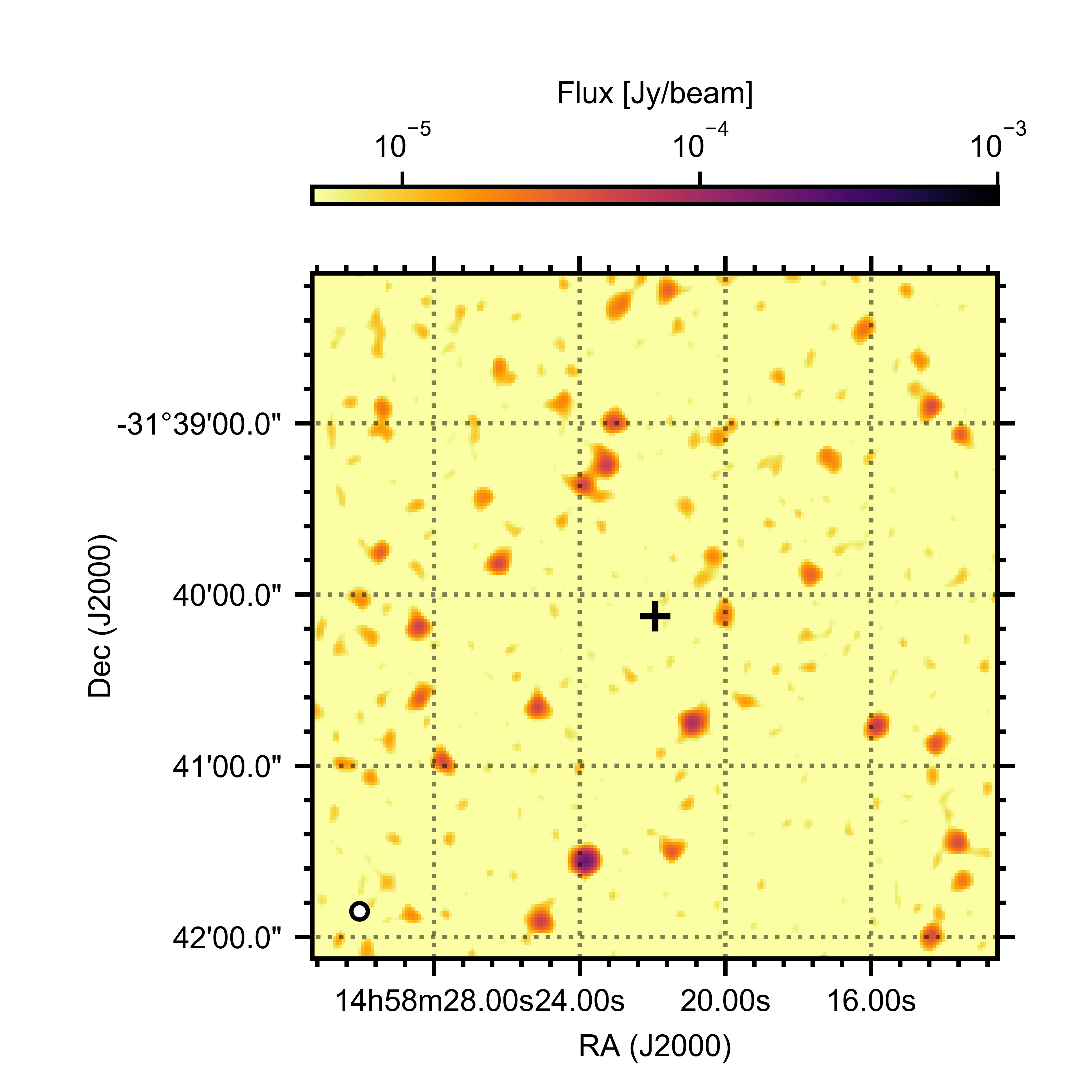}
    \caption{Zoom of the $4\times4$-arcminute field around the position of Cen~X--4 in the MeerKAT field shown in Figure \ref{fig:image_largeFOV}. The beam is shown in the bottom left corner. No point source or extended emission is visible at the position of Cen~X--4, shown by the black cross, at an RMS sensitivity of $4.4$ $\mu$Jy/beam.}
    \label{fig:image_zoom}
\end{figure}

No significant radio emission is detected from the position of Cen~X--4 in any of the four MeerKAT observations. In Figure \ref{fig:image_largeFOV}, we show the large-scale field surrounding Cen~X--4 ($1.4\times1.4$ degree), as observed in the deep, $4.5$-hour observation. The black cross indicates the position of the X-ray binary. While a large number of unresolved point sources and extended (background) sources are visible in the image, the zoom inset displayed in Figure \ref{fig:image_zoom} ($4\times4$ arcminutes) confirms that no radio emission associated with Cen~X--4 is detected. In the deep, $4.5$ hour observation, this non-detection implies a $3$-$\sigma$ radio flux density upper limit of $13$ $\mu$Jy at $1.3$ GHz, as determined from the image RMS calculated over the target position. In the three shorter observations on 7, 9, and 16 January 2021, the non-detection of Cen~X--4 implies $3$-$\sigma$ upper limits of $69$, $54$, and $48$ $\mu$Jy, respectively. These upper limits are plotted in the bottom panels of the light curves in Figure \ref{fig:lightcurve}. For a distance of $1.2$ kpc and assuming a flat radio spectrum, these limits correspond to $5$-GHz radio luminosity limits of $L_R < 1.1\times10^{26}$ erg/s in the deep observation, and $5.9\times10^{26}$ erg/s, $4.7\times10^{26}$ erg/s, and $4.1\times10^{26}$ erg/s in the three shorter observations, respectively.

To test whether short-time-scale flaring is present in the observations, we split the three short observations into two equal halves and re-imaged the field. Again, no radio emission is detected in any image at the position of Cen X-4, at RMS sensitivities a factor $\sqrt{2}$ higher than in the full observations. Similarly, we break up the long quiescent observation into four equal segments lasting two observing scans ($\sim 1$ hour), optimizing the trade-off between time resolution and sensitivity at these low X-ray luminosities. None of the four radio images show significant emission at the target position, at a sensitivity of $9$ $\mu$Jy/bm.

\subsection{X-ray light curves and spectra}

\begin{figure*}
    \centering
    \includegraphics[width=\textwidth]{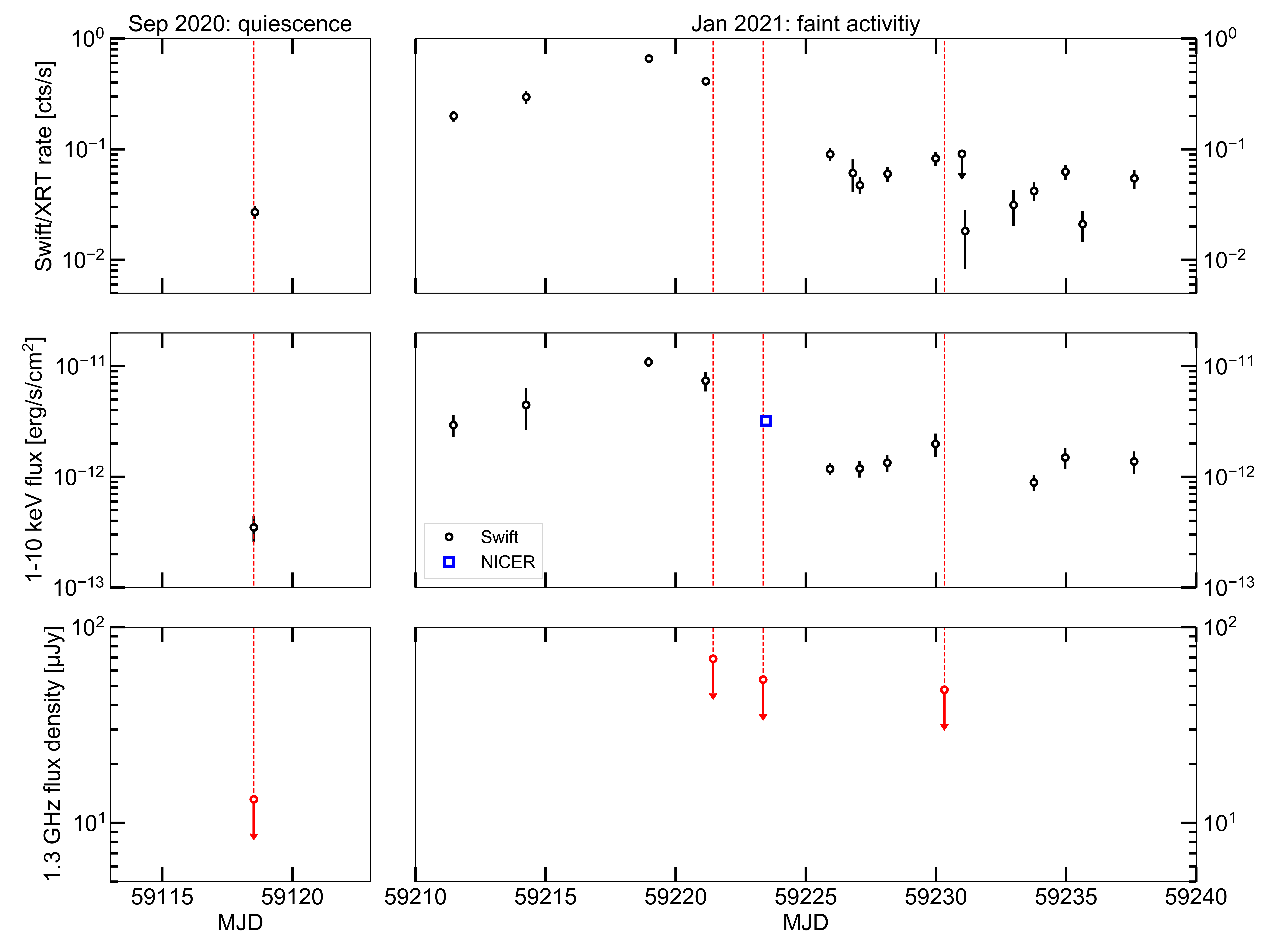}
    \caption{X-ray and radio light curves of Cen~X--4 in September 2020 (left) and January 2021 (right). The top panel shows the \textit{Swift}/XRT count rates, all measured in PC mode. The middle panel shows the $1$--$10$ keV flux measured from the spectral fits to either \textit{Swift} (black circles) or \textit{NICER} (blue square) spectra. The bottom panel shows the radio flux density upper limits for the four MeerKAT observations at $1.3$ GHz.}
    \label{fig:lightcurve}
\end{figure*}

In the top panel of Figure \ref{fig:lightcurve}, we plot the \textit{Swift}/XRT light curve around the deep September 2020 radio and X-ray observations (left) and during the period of faint activity in January 2021 (right). The red dashed lines indicate the times of the four MeerKAT observations. The latter activity shows a rise in X-ray rates over approximately one week, peaking around $0.65$ cts/s in PC mode \citep[see also][]{saikia2021,vandeneijnden2021_atel1}. Cen~X--4 then decayed over a comparable time scale \citep[e.g.,][]{vandeneijnden2021_atel2}, returning to its typical quiescent count rate levels between $\sim 10^{-2}$ and $\sim 10^{-1}$ cts/s \citep{bernardini2013}.

To calculate the X-ray flux from the \textit{Swift} and \textit{NICER} spectra, we followed the approach of \citet{bernardini2013}, who studied the quiescent \textit{Swift}/XRT PC-mode spectra of Cen~X--4 taken daily over a $60$-day period. \citet{bernardini2013} decomposed the spectrum into a soft component emitted by the NS atmosphere and a harder component attributed to the low-level accretion flow. To perform X-ray modelling at the low cumulative number of counts in the \textit{Swift} spectra, they combined observations in three $0.5$--$10$ keV count-rate ranges to derive typical spectral parameters for the atmosphere and accretion-driven emission: low rates below $7\times10^{-2}$ ct/s, medium between $7\times10^{-2}$--$1.1\times10^{-1}$ ct/s, and high above $1.1\times10^{-1}$ ct/s. As shown by Figure \ref{fig:lightcurve}, our observations probe a similar range in X-ray activity of Cen~X--4. As we aim to measure the time-dependent flux, we cannot similarly combine observations. However, we can use the template spectra determined in the above three count rate ranges to infer fluxes, when necessary.

Practically speaking, this means we define three ranges for our \textit{Swift} spectra, each with their own approach: (i) for observations exceeding $100$ source counts, we directly fit the spectral models to the data; (ii) for observations with $<100$ counts and a count rate below $7\times10^{-2}$ ct/s, we fit the overall normalization of the low-rate template spectral model from Bernardini et al. (2013); (iii) for observations with $<100$ counts and a count rate between $7\times10^{-2}$-$1.1\times10^{-1}$ ct/s, we fit the normalization of the medium-rate template spectral model instead. We explain the fits for these three ranges below in more detail. In Table \ref{tab:xrays}, we list the spectral analysis details for each ObsID.

For the three \textit{Swift} spectra with sufficient counts for an actual fit -- case (i) -- we attempted model fits with three models based on \citet{bernardini2013}: a phenomenological, absorbed power law model (\textsc{tbabs*powerlaw}); a physical, absorbed NS atmosphere model (\textsc{tbabs*nsatmos}); and their combination, representing the addition of NS atmospheric and accretion-driven emission (\textsc{tbabs*(nsatmos + powerlaw)}). In all X-ray spectral fits, we fixed the absorption column to $N_H = 8\times10^{20}$ cm$^{-2}$. In all three cases of the high-rate regime, when comparing the single-component models, we find that the power law model provides a statistically better fit than the NS atmosphere model (i.e. a lower C statistic for the same number of free parameters). The composite model does not provide a statistically better fit in any of the spectra, given the C-statistic improvement for the two extra free parameters. To mirror the approach in cases (ii) and (iii) (see below), we use the $1$--$10$ keV unabsorbed flux from this composite model. However, this flux is consistent with the power-law-only flux in all three cases. All fitted parameters, fit statistics, and fluxes are listed in Table \ref{tab:xrays}. We also confirmed that the fitted parameters and measured fluxes were consistent between the \textsc{run\_xrtpipe.pl} and online-pipeline data reduction.

In cases (ii) and (iii), the template model is defined as the third model above, i.e. \textsc{tbabs*(nsatmos + power law)}. The difference between the low and medium rate case lies in the parameters of the NS atmosphere and the relative contributions of the two spectral components. In practice, we took the parameters from Table 4 in \citet{bernardini2013}, converted the $kT^{\infty}$ values to units of Kelvin and the thermal fraction to an \textsc{nsatmos} normalization, and defined a \textsc{constant*tbabs*(nsatmos + powerlaw)} model in \textsc{xspec}. We then fitted the constant for each considered spectrum, keeping all other parameters frozen to their template values. Finally, we calculated the X-ray flux by multiplying the $1$--$10$ keV unabsorbed model fluxes of $1.31\times10^{-12}$ erg/cm$^2$/s (low rates) or $1.60\times10^{-12}$ erg/cm$^2$/s (medium rates) with the fitted constant. We finally calculated the flux errors by propagating the $1$-$\sigma$ error on the constant to the flux. Our approach implies that we use the full X-ray flux; in the Discussion we will explicitly discuss the contributions of the different spectral components to the flux.

Finally, we turned to the single \textit{NICER} spectrum. Cen~X--4 is only detected above the background between $0.5$ and $1.5$ keV. We assume the medium-rate template spectrum from \citet{bernardini2013} and again fit the constant scaling factor. This spectral shape gives a significantly better fit than the low-rate template. Moreover, it fits with the rates in the surrounding \textit{Swift} observations, which exceed the maximum rate for the low-rate template.

The light curve of the $1$--$10$ keV unabsorbed X-ray flux is plotted in the middle panels of Figure \ref{fig:lightcurve}. The evolution clearly follows the count rate profile, although minor differences are visible. For instance, due to differences in spectral shape, the X-ray flux during the September 2020 observation is significantly lower than that after the return to quiescence in January 2021, despite both sharing similar count rates. Also, five count-rate data points do not have an associated X-ray flux due to the limited number of source counts. Finally, the \textit{NICER} flux fits well with the general decay trend during that phase of the light curve, despite the small energy range where Cen~X--4 is detected.

\begin{table*}
\caption{Details of the X-ray flux determination. For the first three ObsIDs, we list the fitted spectral parameters for the three attempted models: \textsc{tbabs*powerlaw}, \textsc{tbabs*nsatmos}, and \textsc{tbabs*(nsatmos + powerlaw)}, per respective row. For the other ten observations, we instead list the measured \textit{Swift}/XRT PC-mode count rate, the adopted template model shape, and the measured model normalization. For all observations, we list the MJD, C-statistic, number of degrees of freedom (dof), and unabsorbed 1--10 keV X-ray flux. In all fits, we fixed interstellar absorption to $N_H = 8\times10^{20}$ cm$^{-2}$. Due to a low total number of source counts, we did not perform a spectral analysis for \textit{Swift} ObsIDs 00035324073, 00035324077, 00035324078, 00035324079, and 00035324083. All \textit{Swift}/XRT count rates in the top panels of Figure \ref{fig:lightcurve} are available via the link provided in the Data Availability statement. *Parameter pegged at lower limit.}
\label{tab:xrays}
\begin{tabular}{lllllllll}
\hline
Obs & ObsID & MJD & $\log T_{\rm eff}$ & $N_{\rm nsa}$ & $\Gamma$ & $N_{\rm po}$ & C (dof) & $1$--$10$ keV flux \\
& & & [K] & & & [keV$^{-1}$cm$^{-2}$s$^{-1}$] & & [erg cm$^{-2}$s$^{-1}$] \\ \hline
\textit{Swift} & 00088937006 & 59118.52 & -- & -- & $3.0\pm0.3$ & $(2.3\pm0.3)\times10^{-4}$ & 40.2 (93) & $(3.3 \pm 0.9)\times10^{-13}$ \\
& & & $6.3\pm0.1$ & $\left(2.1^{+3.4}_{-1.1}\right)\times10^{-2}$ & -- & -- & 49.8 (93) & $(2.8 \pm 0.6)\times10^{-13}$ \\
 & & & 5.0* & $\geq 0$* & $2.8\pm0.4$ & $\left(2.1^{+0.4}_{-0.6}\right)\times10^{-4}$ & 39.5 (91) & $(3.5 \pm 1.0)\times10^{-13}$ \\ \hline
\textit{Swift} & 00035324070 & 59218.96 & -- & -- & $2.4\pm0.1$ & $(4.4\pm0.2)\times10^{-3}$ & 195.9 (189) & $(1.0 \pm 0.1)\times10^{-11}$ \\
& & & $>6.47$ & $\left(5.8^{+3.6}_{-0.4}\right)\times10^{-2}$ & -- & -- & 268.2 (189) & $(8.9\pm0.6)\times10^{-12}$ \\
 & & & $6.25^{+0.11}_{-0.07}$ & $\left(3.2^{+3.6}_{-2.6}\right)\times10^{-1}$ & $1.9^{+0.4}_{-0.6}$ & $(2.0\pm0.1)\times10^{-3}$ & 190.6 (187) & $(1.1 \pm 0.1)\times10^{-11}$ \\ \hline
\textit{Swift} & 00035324071 & 59221.15 & -- & -- & $2.7\pm0.2$ & $(3.4\pm0.3)\times10^{-3}$ & 86.0 (89) & $(6.4\pm0.9)\times10^{-12}$\\
& & & $6.45\pm0.04$ & $\left(6.2^{+3.0}_{-2.0}\right)\times10^{-2}$ & -- & -- &  132.3 (89) & $(6.0\pm0.7)\times10^{-12}$ \\
& & & $6.0\pm0.1$ & $3.6^{+7.2}_{-2.2}$ & $1.6\pm0.6$ & $\left(9.1^{+1.1}_{-0.6}\right)\times10^{-4}$ & 82.3 (87) & $(7.4\pm1.5)\times10^{-12}$ \\
\hline\hline
Obs & ObsID & MJD & PC Rate & Template & \multicolumn{2}{l}{Template} & C (dof) & $1$--$10$ keV flux \\
& & & [cts s$^{-1}$] & spectrum & \multicolumn{2}{l}{normalization} & & [erg s$^{-1}$cm$^{-2}$] \\ \hline
\textit{Swift} & 00035324068 & 59211.46 & $1.1\times10^{-1}$ & Medium & $1.84 \pm 0.22$ & & 61.4 (64) & $(2.9 \pm 0.6)\times10^{-12}$ \\
\textit{Swift} & 00035324069 & 59214.25 & $1.1\times10^{-1}$ & Medium & $2.79 \pm 0.41$ & & 42.9 (41) & $(4.5 \pm 1.8)\times10^{-12}$ \\
\textit{NICER} & 3652010901 & 59223.46 & N/A & Medium & $2.07 \pm 0.06$ & & 102.1 (63) & $(3.2 \pm 0.2)\times10^{-12}$ \\
\textit{Swift} & 00035324072 & 59225.93 & $7.3\times10^{-2}$ & Medium & $0.74 \pm 0.12$ & & 31.0 (35) & $(1.2 \pm 0.1)\times10^{-12}$ \\
\textit{Swift} & 00035324074 & 59227.06 & $4.2\times10^{-2}$ & Low & $0.91 \pm 0.17$ & & 29.4 (30) & $(1.2 \pm 0.1)\times10^{-12}$ \\
\textit{Swift} & 00035324075 & 59228.13 & $5.1\times10^{-2}$ & Low & $1.03 \pm 0.18$ & & 22.4 (35) & $(1.3 \pm 0.2)\times10^{-12}$ \\
\textit{Swift} & 00035324076 & 59229.98 & $6.0\times10^{-2}$ & Low & $1.52 \pm 0.25$ & & 38.5 (34) & $(2.0 \pm 0.5)\times10^{-12}$ \\
\textit{Swift} & 00035324081 & 59233.76 & $3.6\times10^{-2}$ & Low & $0.68 \pm 0.17$ & & 21.0 (22) & $(8.9 \pm 0.2)\times10^{-13}$ \\
\textit{Swift} & 00035324082 & 59234.96 & $4.8\times10^{-2}$ & Low & $1.15 \pm 0.21$ & & 27.2 (34) & $(1.5 \pm 0.3)\times10^{-12}$ \\
\textit{Swift} & 00035324084 & 59237.61 & $3.8\times10^{-2}$ & Low & $1.06 \pm 0.23$ & & 26.0 (23) & $(1.4 \pm 0.3)\times10^{-12}$ \\ \hline
\end{tabular}\\
\end{table*}

\subsection{The X-ray -- radio luminosity plane}

To further investigate the low-luminosity inflow/outflow coupling in Cen~X--4, we combine X-ray and radio observations and place them on the X-ray--radio luminosity diagram. We match up each radio observation with the X-ray observation taken within $24$ hours (ObsIDs 00088937006, 00035324071, 3652010901, and 00035324076, chronologically). To convert the X-ray fluxes to $1$--$10$ keV luminosities, we assume a distance of $1.2$ kpc \citep[][although see Section \ref{sec:disc} for a discussion on the distance]{chevalier1989}. To calculate the radio luminosity upper limits from the flux densities, we calculate $L_R = \nu L_\nu$ at a frequency of 5 GHz. This X-ray energy band and radio frequency are chosen to be the same as in the X-ray binary $L_X$--$L_R$ catalogue by \citet{bahramian2018b}.

\begin{figure*}
    \centering
    \includegraphics[width=\textwidth]{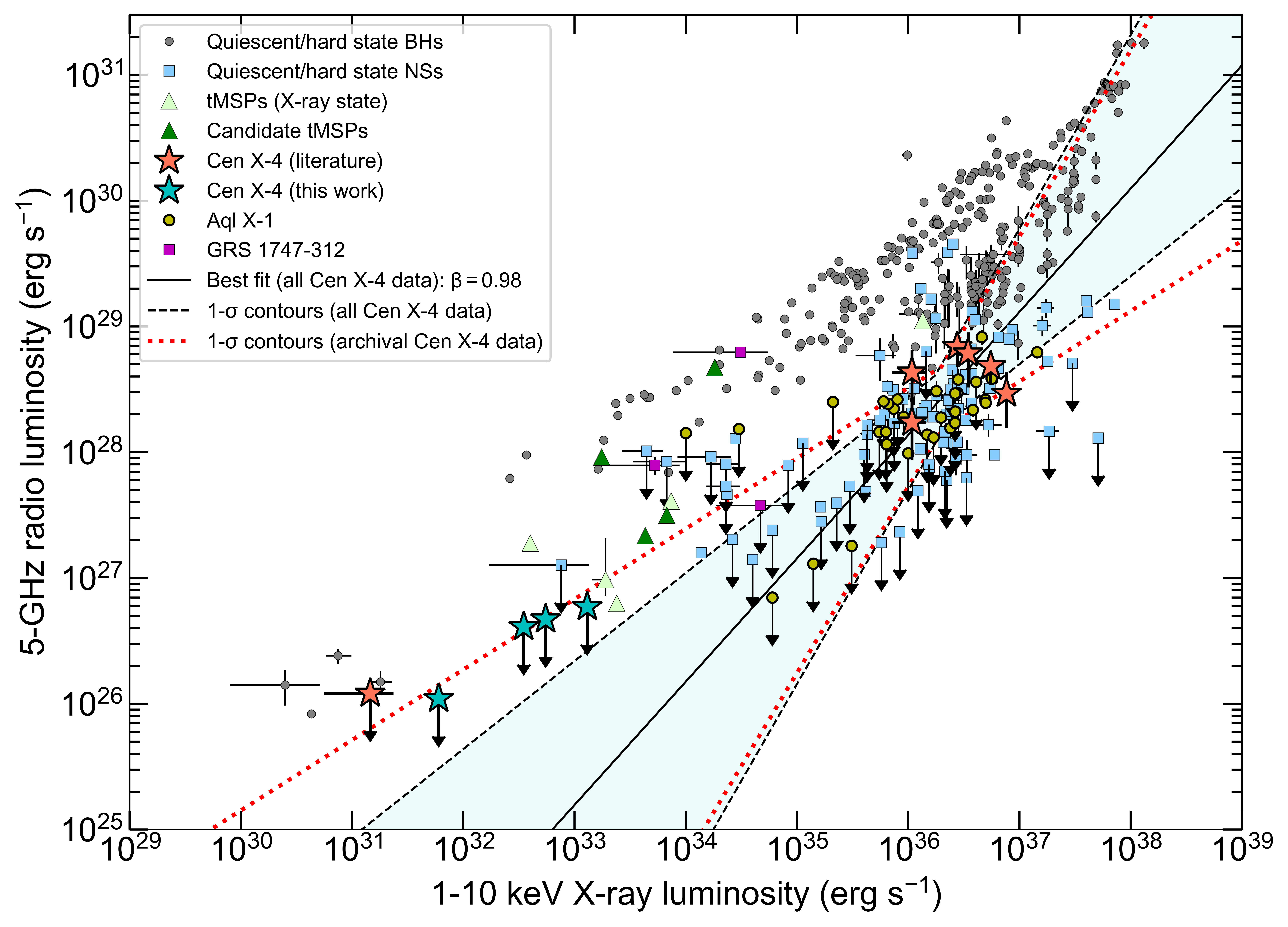}
    \caption{The $1$--$10$ keV X-ray -- 5-GHz radio luminosity diagram for low-mass X-ray binaries. The grey circles represent hard-state black hole systems. Light blue squares show neutron stars, particularly hard-state atoll sources and AMXPs. tMSPs are shown as the light green upwards pointing triangles, while four tMSPs are shown as the upwards dark green triangles. We also highlight several sources in particular: Cen~X--4 as the stars (red for archival data, turquoise for data from this work), Aql X-1 as the yellow circles, and GRS 1747-312 as the purple squares. For Aql X-1, the compilation of data is taken from \citet{gusinskaia20}, where we only plot points with an associated hardness ratio exceeding 0.75. The black line indicates the best \textsc{linmix} fit to all Cen~X--4 data, with the shaded area showing the $1$-$\sigma$ confidence region. The red dotted lines indicate the  $1$-$\sigma$ confidence region without the new MeerKAT observations. Plot based on the catalogue by \citet{bahramian2018b} with more recent data added (see text).}
    \label{fig:lxlr}
\end{figure*}

In Figure \ref{fig:lxlr}, we plot the X-ray--radio luminosity diagram for low-mass X-ray binary systems, based largely on \citet{bahramian2018b}. The green circles show hard-state BH systems, while NSs in equivalent states\footnote{More specifically, atolls in the island state and the Accreting Millisecond X-ray Pulsars.} are shown in blue squares. The transitional millisecond pulsars (tMSPs) in their X-ray bright state are shown as the light-green upwards triangles. A number of NS systems are specifically highlighted. Cen~X--4 in shown by the stars, where red stars show archival data from \citet{tudor17}, \citet{hjellming1988}, and \citet{kaluzienski1980}, and turquoise stars show the four observations presented in this work. Aql X-1 is shown with yellow circles, taken from the detailed study by \citet{gusinskaia20}, which is the most commonly observed system of the NSs plotted here (note that we only plot points where \citealt{gusinskaia20} calculate a hardness ratio of at least 0.75). Three observations of GRS 1747-312, that we will explicitly discuss in Section \ref{sec:disc}, are shown by the magenta squares \citep{panurach2021}. Using distances from \citet{tremou2018}, we also include the quasi-simultaneous data from \citet{panurach2021} for the globular cluster sources 4U 1746-37, XB 1832-330, X 1850-087, and M15 X-3, which were all observed when actively accreting, although they were relatively X-ray faint. 

The deepest MeerKAT radio upper limit of Cen~X--4 is comparable to the deep Very Large Array limit obtained by \citet{tudor17}, but at a $\sim$four times higher X-ray luminosity. As a result, the new non-detection reported here is more constraining to the low-luminosity end of the X-ray--radio luminosity correlation for the source. The radio upper limits from the further three MeerKAT observations are at higher luminosities, but as they are taken during the period of faint X-ray activity, they may similarly constrain the slope of this correlation. More generally, radio detections of confirmed NS below $L_X = 10^{34}$ erg/s are rare. Our observations add four more points to this poorly-explored regime, offering a particularly interesting comparison to the tMSPs: this source class has been radio-detected at similar X-ray luminosities, with radio luminosities above or close to the detection threshold of our Cen~X--4 observations (see Discussion).

To assess the effect of our new observations on the inferred $L_X$--$L_R$ coupling, we follow the \textsc{linmix}-approach, originally developed by \citep{kelly2007}. This MCMC-method fits a linear model to data with errors in both the dependent and independent variable, fully accounting for upper limits in the former. It was first applied to the study of the X-ray binary $L_X$--$L_R$ plane by \citet{gallo2014}, and further adapted by \citet{gusinskaia20} and Van den Eijnden et al. (submitted) to estimate the uncertainties and account for distance errors. To apply this method, we linearize the correlation model $L_R/L_{R,0} = \xi (L_X/L_{X,0})^\beta$ to
\begin{equation}
    \log L_R - \log L_{R,0} = \log \xi + \beta \left(\log L_X - \log L_{X,0}\right)  \text{ .}
\end{equation}

We then apply the \textsc{python}-implementation of the method\footnote{Available via \url{https://github.com/jmeyers314/linmix}} to fit the offset and slope of this linear equation, as well as the data's intrinsic scatter, parameterized by a Gaussian standard deviation $\sigma$. In particular, we determine for each parameter the median and $16^{\rm th}$/$84^{\rm th}$ percentile from $10^4$ draws of the posterior distribution. We then follow \citet{gusinskaia20} and Van den Eijnden et al. (submitted) and repeat this approach $500$ times, each time re-drawing the assumed distance (i.e. scaling the fitted luminosities by a factor $(D_i/D)^2$ for iteration i) from a Gaussian distribution centred at $1.2$ kpc with a $0.3$ kpc standard deviation. We then determine the fitted parameters and their $1$-sigma uncertainties as the mean value of the $500$ saved medians and $16^{\rm th}$/$84^{\rm th}$ percentiles, respectively.

In Figure \ref{fig:lxlr}, the black line shows the best fit to all Cen~X--4 observations, with $\log \xi = -0.14^{+0.22}_{-0.27}$ and $\beta = 0.98^{+0.60}_{-0.28}$. The lightly-shaded area indicates the $1$-$\sigma$ confidence interval of the fit\footnote{We note that combining the two X-ray brightest radio epochs, which also returns a non-detection, does not lead to significantly different results of the fit.}. To compare these values to the archival data from \citet{tudor17}, we exactly repeat these fits without the four MeerKAT data points. In that case, we find $\log \xi_{\rm archival} = -0.12^{+0.23}_{-0.27}$ and $\beta_{\rm archival} = 0.84^{+0.64}_{-0.28}$, which set the $1$-$\sigma$ region enclosed by the red dotted lines in Figure \ref{fig:lxlr}. Expectedly, the fitted normalisations are consistent, as these are most strongly constrained by the radio detections during outburst. Similarly, the upper limits on the slope $\beta$ are almost identical, as these are not affected by the addition of radio upper limits at low $L_X$. The minimum slope is, on the other hand, more strongly constrained. This can be deduced from comparing the $1$-$\sigma$ regions in Figure \ref{fig:lxlr} at low $L_X$. Alternatively, we can compare the $90$\% lower limit on $\beta$ from our fit: this limit increases from $\beta \geq 0.51$\footnote{This value is consistent with the lower limit reported by \citet{tudor17}, i.e. $\beta > 0.5$.} to $\beta \geq 0.66$ with the addition of the MeerKAT observations. 

\section{Discussion}
\label{sec:disc}

In this paper, we have presented four MeerKAT radio observations of the NS LMXB Cen~X--4 during quiescence and a period of faint X-ray activity. Cen~X--4 is not detected in any of the radio observations, providing deep limits on the correlation between X-ray and radio luminosity in the low-luminosity regime (e.g. $L_X \lesssim 10^{33}$ erg/s). Here, we will discuss the implication of these measurements for our understanding of (different types of) NSs in the $L_X$--$L_R$ plane, the origin of the measured X-ray luminosity, and the prospects for low-luminosity NS LMXB radio studies using more sensitive future arrays.

\subsection{A note on the distance to Cen X-4}

In our analysis, we assumed the $1.2\pm0.3$ kpc distance measured by \citet{chevalier1989} to convert fluxes and flux densities to luminosities. Recent \textit{Gaia} parallax measurements \citep{gaia_edr3}, however, imply a slightly larger distance: at $68$\% confidence, converting the parallax to a distance using the prior for Galactic X-ray binaries from \citet{atri2019} and applying zero-point corrections, we find $d = 1.87^{+0.75}_{-0.42}$ kpc. To assess the effect of this larger distance on our inferences, we repeat our fits in the $L_X$--$L_R$ plane after correcting all luminosities by a factor $(1.87/1.2)^2$. This calculation returns consistent parameters: we find $\beta_{\rm gaia} = 0.98^{+0.58}_{-0.28}$ and $\log \xi_{\rm gaia} = -0.13^{+0.23}_{-0.38}$. Finding the same slope is unsurprising, given that both $L_X$ and $L_R$ scale in the same manner with distance. Finding a consistent normalization at $L_{X,0}$ occurs because the best-fit slope is consistent with linear. Therefore, the main effect of the larger \textit{Gaia} distance is in the comparison with other sources. We will therefore note, in the below discussions, where this may play a role.

\subsection{The origin of the X-ray emission}

In this work, we have followed the common approach to the X-ray spectra, where the unabsorbed X-ray fluxes from the entire spectral model are calculated to obtain the $L_X$ measurements. However, at the low X-ray luminosities of Cen~X--4, the soft X-rays are fitted by a NS atmosphere model, while only the harder X-rays are dominated by the fitted power law component. When our \textit{Swift} spectra contain enough counts for a model fit, we find that this two-component model does not fit better than a power-law-only model. However, this could likely be caused by the low signal-to-noise, given that analyses by \citet{bernardini2013} and \citet{chakrabarty14} find significant soft thermal emission at similar X-ray luminosity in higher quality spectra. For this reason, and for consistency with the lower quality spectra analysed using two-component template spectral models, we used the two-component X-ray fluxes for all observations.

Since $L_X$--$L_R$ correlations are interpreted as the observational signature of a coupling between inflow, i.e. mass accretion rate, and outflow, i.e. jet power, we can wonder whether the two-component X-ray flux correctly traces the mass accretion rate. The detailed modeling of the high-energy component by \citet{chakrabarty14} is consistent with this component originating in a radiatively-inefficient accretion flow, while \citet{bernardini2013} similarly infer that the power-law-emission seen in \textit{Swift} spectra likely originates from the accretion flow at low mass accretion rate. Moreover, \citet{bernardini2013} find that the thermal X-rays from the NS atmosphere and the harder X-ray component are variable in a coupled fashion, leading to their conclusion that the full X-ray spectrum is powered by the accretion flow: either directly in the hard component or indirectly, after accreting onto the neutron star surface, in the soft, thermal component. In that interpretation, the combined X-ray luminosity can indeed be used in our $L_X$--$L_R$ modelling.

For a scenario where all components of the X-ray spectrum are directly or indirectly powered by accretion (instead of e.g. the neutron star surface cooling from previous activity but not actively heated by the instantaneous accretion), the correlation between accretion rate and X-ray luminosity is expected to be linear: as advection of energy across an event horizon is not possible, the process is ultimately expected to be radiatively efficient. In that scenario, although the relative contributions of different spectral components may differ as a function of accretion rate, the correlation between $L_X$ and $L_R$ would remain the same from quiescence to the outburst peak. The lack of radio detections of NS LMXBs below $L_X = 10^{34}$ erg/s has made it challenging to test this hypothesis; a recent analysis at higher X-ray luminosities for the NS LMXB Aql X-1 by \citet{fijma2022} did not reveal evidence for any significant effects of the relative contributions of spectral components on the coupling. However, that study also showed how signal-to-noise with current facilities, at distances of several kpc, makes it hard to disentangle such X-ray components, and how the presence of accretion state changes dominates the changes in radio behaviour. Therefore, observations covering quiescence and the rise of the outburst may be most suitable for such a study. However, our MeerKAT results show that current sensitivities in monitoring observations are insufficient for low-$L_X$ radio detections, even at $1.2$ kpc.

\subsection{NS LMXB jets at low X-ray luminosity}

\subsubsection{Phenomenological source comparisons}

With our radio limits at low X-ray luminosity, we constrain the slope of the $L_X$--$L_R$ correlation for Cen~X--4 to $\beta = 0.98^{+0.59}_{-0.28}$. Using the same fitting approach, \citet{gallo18} found a correlation slope of $\beta = 0.44^{+0.05}_{-0.04}$ for the entire sample of NS LMXBs, significantly shallower than our Cen~X--4 result. As Cen~X--4 is likely an atoll source, a more apt comparison, however, may be with the atoll-only fit performed by \citet{gallo18}. This atoll-only correlation slope, $\beta = 0.71^{+0.11}_{-0.09}$ is consistent within its $1$-$\sigma$ errors with Cen~X--4. As discussed in \citet{gallo18}, the measured slopes may be affected by the X-ray luminosity distribution of the fitted observations: for the total sample of atolls, which are predominantly observed (and especially radio detected) above $10^{36}$ erg/s, this may lead to shallower slopes. With the largest range in X-ray luminosity of any (likely) atoll source, we expect that such issues do not affect our Cen~X--4 measurement to the same extent. Recently, \citet{gusinskaia20} presented a detailed look at the behaviour of Aql X-1 in the $L_X$--$L_R$ plane, across various outbursts. When, for the first time, including deep radio limits (i.e. $L_R \lesssim 10^{27}$ erg/s) between $L_X = 6\times10^{34}$ and $3\times10^{35}$ erg/s, they find a correlation slope of $\beta = 1.17^{+0.30}_{-0.21}$ -- again consistent with our Cen~X--4 measurement.

Moving beyond comparisons with atoll source, we can turn to other sub-classes of accreting NSs. For instance, our Cen~X--4 radio non-detections during its period of faint activity provide stricter confirmation that tMSPs are radio-brighter than other NSs at these low X-ray luminosities (note that the same does not necessarily hold above, e.g., $L_X = 10^{35}$ erg/s; \citealt{russell18}; \citealt{vandeneijnden2018_igr17379}). In particular, the observations of the first-discovered tMSP, PSR J1023+0038, by \citet{deller2015} lie above our radio upper limits and are excluded from the $1$-$\sigma$ region for Cen~X--4's $L_X$--$L_R$ correlation, for the distance of $1.2\pm0.3$ kpc. This finding confirms the suspicion from \citet{tudor17} that observations at higher $L_X$ than theirs, would either detect or disprove the formation of jets in Cen~X--4 as radio-bright as tMSPs \citep[although we note that a jet origin for the radio emission of tMSPs is not confirmed; e.g.][]{bogdanov18}. 

In Figure \ref{fig:lxlr}, we also show four candidate but unconfirmed tMSPs that were recently studied in the $L_X$--$L_R$ plane as the dark green upward triangles: 3FGL J0427.9-6704 \citep{li2020}, 3FGL J1544.6-1125 \citep{jaodand2021}, NGC 6652B \citep{paduano2021}, and CXOU J110926.4-650224 \citep[][assuming a $4$ kpc distance]{cotizelati2021}. We plot one point per source: the radio brightest of four observations for 3FGL J1544.6-1125, and the average X-ray and radio luminosity of the other three. All four are also detected as radio-bright systems, sometimes even consistent with the BH population in the $L_X$--$L_R$ diagram. Those observations, if the systems are indeed confirmed to be tMSPs, further indicate a clear difference with Cen~X--4 at low $L_X$. If the relative radio brightness of tMSPs is related to interactions between the low-level accretion flow and the pulsar magnetosphere, those interactions should be markedly different than in non-transitioning LMXBs.

Finally, we note that the recent work by \citet{panurach2021}, studying NS LMXBs in globular clusters with the MAVERIC survey, has revealed that low-level-accreting neutron stars can be radio variable. In particular, the transient NS LMXB GRS 1747-312, shown in Figure \ref{fig:lxlr} as the magenta squares, shows a mixture of radio detections and non-detections, implying radio variability by more than an order of magnitude despite a relatively small range in X-ray luminosity ($5\times10^{33}$--$5\times10^{34}$ erg/s). At a slightly higher $L_X \approx 10^{36}$ erg/s, \citet{panurach2021} find that the persistent NS LMXB X 1850-087 is similarly radio variable. In the tMSP population, PSR J1023+0038 also famously shows strong radio variability during its X-ray-bright state: highly-structured moding behaviour, anti-correlated with the X-ray band, as well as erratic radio flaring, both on time scales of minutes \citep[e.g.][]{bogdanov18}. During the faint activity period of Cen X-4, we do not find evidence for variability during our observations, when splitting the observing block in two. Longer observations during faint periods of accretion may, in the future, provide better constraints on such variability. Similarly, the longer MeerKAT observation did not reveal evidence for radio variability, during the source's quiescent state.

\subsubsection{Correlation breaks and radio-dark propeller outflows}

The phenomenological measurements of the $L_X$--$L_R$ relation discussed in the previous section (i.e. for Aql X-1, all atolls, and ours for Cen~X--4), fundamentally assume a single correlation to hold between quiescence and the peak of the outburst (or the state transition where the source may show radio quenching). \citet{gusinskaia20} instead also discuss an alternative scenario, where the $L_X$--$L_R$ correlation of Aql X-1 shows a sharp cutoff a X-ray luminosities below $\sim 5\times10^{35}$ erg/s\footnote{We note that \citep{plotkin2017} rule out a similar cutoff in the X-ray -- radio luminosity relation for the black hole LMXB V404 Cyg. In that work, such a cutoff would be associated with the jet dominating the X-ray emission a low accretion rate.}. Generally, NS LMXBs appear to change from regular radio detections above $L_X \approx 10^{36}$ erg/s to mostly radio non-detections at lower X-ray luminosities. While this change is most simplistically explained via a relatively steep, single $L_X$--$L_R$ correlation, one can explore an explanation for a break in the correlation as well. For Cen~X--4 (or any other source, for that matter), current data do not allow to distinguish between these scenarios: this is particularly difficult due to the $\sim 3$ order of magnitude in X-ray luminosity that has never been covered with radio observations. Similarly, if such a break in the correlation would exist, current data do not constrain whether it occurs at the same X-ray luminosity in different sources.

The X-ray luminosity range where radio non-detections become dominant, is of similar order of magnitude as the range where the onset of the propeller regime is expected. In this regime, the magnetospheric radius, where the NS magnetic field and accretion disk pressure are in balance, moves outside the co-rotation radius due to a decrease in the accretion rate. At the co-rotation radius, the NS spin equals the Keplerian frequency of the disk. Therefore, when the magnetospheric radius moves beyond the co-rotation radius, a centrifugal barrier may be created, halting the accretion flow. The accretion luminosity where this is expected to occur can be written as

\begin{equation}
L_{\rm prop} = 1.6\times10^{33} \text{ } B_8^2 \text{ } \nu_{100}^{7/3} \text{ erg/s,}
\end{equation}
where we assume disk accretion (instead of spherical/wind accretion), $B_8$ is the magnetic field divided by $10^8$ G, and $\nu_{100}$ is the spin frequency divided by $100$ Hz \citep{tsygankov17}. For e.g, Aql X-1, with its $550$ Hz spin, $L_{\rm prop}$ can easily match $\sim 5\times10^{35}$ erg/s for a realistic magnetic field of a few times $10^8$ G; more generally, for AMXPs with spins at hundreds of Hz, $L_{\rm prop}$ can reasonably be expected to lie between $10^{35}$ and $10^{36}$ erg/s. In the propeller scenario, decaying below $L_{\rm prop}$ would expel material as angular momentum is transferred to the accreting material \citep{illarionov75}. Alternatively, in other models and circumstances, the disk may instead be trapped around the magnetospheric radius, preventing efficient accretion \citep{spruit93,dangelo10}.

The onset of the propeller regime has regularly been suggested to be associated with enhanced radio emission from the outflow. For instance, to explain the difference between the radio luminosity during the decay of outbursts of the AMXP SAX J1808.4-3658, \citet{tudor17} invoke a propeller-driven outflow to explain radio detections below $L_X = 10^{36}$ erg/s. However, one may wonder if additional radio emission, or any radio emission, is always expected in this regime. For instance, if the disk is trapped (mostly expected when $L_X$ drops slightly below $L_{\rm prop}$), it may prevent effective jet launching altogether. Moreover, simulations of the propeller regime by \citet{ustyugova2006}, \citet{romanova09}, and \citet{lii2014} show the onset of a two-component outflow: a fast and collimated outflow combined with a more massive, wider-angled and slower wind-like outflow. While radio emission may be expected from the former, it is unclear where the energetic balance between the outflows lies. If most of the energy is carried away by the wind-like slow outflow, which is not expected to emit brightly at radio frequencies, the radio luminosity of NS LMXBs may be suppressed when decreasing below $L_{\rm prop}$.

We reiterate that the onset of such a `radio-dark' propeller outflow is not necessary to explain current observations; as stated earlier, a relatively steep, single $L_X$--$L_R$ correlation (per source or the full sample) can similarly account for the currently available data. As discussed below, more sensitive and regular radio monitoring of this low-luminosity regime could provide the data to observationally test the single-correlation hypothesis. However, two brief points are worth stating: firstly, if a steepening/breaking of the $L_X$--$L_R$ correlation would be observed in future campaigns, the explanation will likely be related to the NS properties, given the lack of such steepening in black hole systems. Secondly, the idea of the propeller regime suppressing the radio luminosity in an X-ray binary was speculated on earlier by \citet{vandeneijnden2019_reb} for strongly-magnetic but slowly-spinning accreting NSs (i.e. $B \gtrsim 10^{12}$ G, $\nu < 1$ Hz). In particular, in Swift J0243.6+6124, the archetypal example of a transient, radio-detected X-ray binary in that class, radio emission is observed to turn on and off rapidly across a narrow range in X-ray luminosity. This luminosity could be associated with the propeller transition for reasonable magnetic field strengths for its NS, which are similar to the recent cyclotron line estimates of its magnetic field \citep{kong2022}.

\subsubsection{Observational tests with future observatories}

With the planned sensitivity of future arrays, such as the next-generation VLA (ngVLA) or the Square Kilometer Array (SKA, specifically SKA-Mid), we will be able to probe at least an order of magnitude deeper to search for quiescent radio emission from Cen~X--4. For instance, the ngVLA's intended $3$-$\sigma$ detection limit at a distance of $1.2$ kpc in a short, $15$-minute observation is $1.2\times10^{25}$ erg/s \citep{selina18}. At these sensitivities, it is possible to directly test for the presence of a break in the $L_X$--$L_R$ correlation at low X-ray luminosities, and therefore test the idea of dark propeller outflows. If such a correlation break is not observed, it can instead probe the radiative efficiency of the accretion flow down to low accretion rate by measuring or constraining the correlation slope further: a single radio non-detection at $L_X = 10^{32}$ erg/s, with the $15$-minute ngVLA sensitivity, would constrain the slope to $\beta \geq 0.8$ ($1$-$\sigma$ limit). Taking this further, not detecting Cen~X--4 during an exact repeat of our MeerKAT campaign --- i.e. at the same X-ray luminosities but with the $15$-minute ngVLA sensitivity --- will imply that $\beta \geq 1.1-1.2$: a limit that starts to constrain the accretion flow to be radiatively efficient \citep{migliari06}. 

On a more systematic level, these future radio sensitivity levels could allow for systematic monitoring of quiescent (NS) LMXBs to detect any low-level activity at larger distances than that of Cen~X--4. Such systematic programs could (i) detect more examples of low-level activity (i.e. $L_X < 10^{34}$ erg/s) associated with mis-fired outbursts, and therefore reveal how common such mis-fired outbursts are; (ii) detect the early onset of outbursts to allow for detailed $L_X$--$L_R$ monitoring; or (iii) provide deep combined images of the quiescent LMXB if no activity is detected. The low-level activity in Cen~X--4 was identified by \citet{waterval2020} due to its small distance in combination with long-term optical monitoring by the XB-NEWS project \citep{xbnews}. Currently, no complementary program to XB-NEWS exists at other wavelengths, and the low-level activity falls below the sensitivity of current and planned all-sky X-ray monitors. The ngVLA and SKA-Mid would be suitable for such complementary low-level activity programs, aimed at the jets instead of the inflow observed in optical: in short, $5$-minute observations with the ngVLA, one reaches $3$-$\sigma$ detection thresholds of $L_R \approx 2\times10^{26}$ erg/s at 4 kpc. A similar program for BH LMXBs may require even shorter observing times or be possible with sub-arrays to increase observing efficiency: the better established $L_X$--$L_R$ relation towards quiescence for BHs also removes uncertainty caused by the uncertain nature of this relation at low $L_X$ for NSs.

\section{Acknowledgments}

The authors thank the anonymous referee for a constructive report. JvdE thanks Joe Bright for discussions on MeerKAT data reduction. JvdE is supported by a Lee Hysan Junior Research Fellowship awarded by St. Hilda's College. We thank the staff at the South African Radio Astronomy Observatory (SARAO) for schedul- ing these observations. The MeerKAT telescope is operated by the South African Radio Astronomy Observatory, which is a facility of the National Research Foundation, an agency of the Department of Science and Innovation. This work was carried out in part using facilities and data processing pipelines developed at the Inter-University Institute for Data Intensive Astronomy (IDIA). IDIA is a partnership of the Universities of Cape Town, of the Western Cape and of Pretoria. The authors acknowledge the use of public data from the Swift data archive. This research has made use of data and software provided by the High Energy Astrophysics Science Archive Research Center (HEASARC) and NASA's Astrophysics Data System Bibliographic Services. This work has made use of data from the European Space Agency (ESA) mission {\it Gaia} (\url{https://www.cosmos.esa.int/gaia}), processed by the {\it Gaia} Data Processing and Analysis Consortium (DPAC, \url{https://www.cosmos.esa.int/web/gaia/dpac/consortium}). Funding for the DPAC has been provided by national institutions, in particular the institutions participating in the {\it Gaia} Multilateral Agreement. TDR acknowledges financial contribution from the agreement ASI-INAF n.2017-14-H.0. GRS is supported by NSERC Discovery Grant RGPIN-2021-04001. This work makes use of several python packages, namely \textsc{numpy} \citep{oliphant_numpy}, \textsc{astropy} \citep{astropy13,astropy18}, \textsc{matplotlib} \citep{hunter07}, and \textsc{aplpy} \citep{robitaille12}.

\section*{Data Availability}

A GitHub repository with a \textsc{jupyter} notebook and all underlying data to reproduce the figures and analysis in this work will be made public upon acceptance and publication via \url{https://github.com/jvandeneijnden/MeerKATCampaignOfCenXm4}. 




\input{main.bbl}



\bsp	
\label{lastpage}
\end{document}